# Habit Coach: Customising RAG-based chatbots to support behavior change

Arian Fooroogh Mand Arabi[1,*], Cansu Koyuturk[1], Michael O'Mahony[2,3], Raffaella Calati[1] and Dimitri Ognibene[1,*]

[1] *Dept. Psychology, Università degli Studi di Milano Bicocca, Milan, Italy*

[2] *School of Computer Science, Technological University Dublin, Ireland*

[3] *SFI Centre for Research Training in Machine Learning at Technological University Dublin, Ireland*

**Abstract**

This paper presents the iterative development of Habit Coach, a GPT-based chatbot designed to support users in habit change through personalized interaction. Employing a user-centered design approach, we developed the chatbot using a Retrieval-Augmented Generation (RAG) system, which enables behavior personalization without retraining the underlying language model (GPT-4). The system leverages document retrieval and specialized prompts to tailor interactions, drawing from Cognitive Behavioral Therapy (CBT) and narrative therapy techniques. A key challenge in the development process was the difficulty of translating declarative knowledge into effective interaction behaviors. In the initial phase, the chatbot was provided with declarative knowledge about CBT via reference textbooks and high-level conversational goals. However, this approach resulted in imprecise and inefficient behavior, as the GPT model struggled to convert static information into dynamic and contextually appropriate interactions. This highlighted the limitations of relying solely on declarative knowledge to guide chatbot behavior, particularly in nuanced, therapeutic conversations. Over four iterations, we addressed this issue by gradually transitioning towards procedural knowledge, refining the chatbot's interaction strategies and improving its overall effectiveness. In the final evaluation, 5 participants engaged with the chatbot over five consecutive days, receiving individualized CBT interventions. The Self-Report Habit Index (SRHI) was used to measure habit strength before and after the intervention, revealing a reduction in habit strength post-intervention. These results underscore the importance of procedural knowledge in driving effective, personalized behavior change support in RAG-based systems.

## 1. Introduction

Advanced capabilities of Large Language Models (LLMs) are gradually changing the landscape of conversational systems through their instruction-following abilities, learning from human feedback, and adjusting the conversational experience to better meet user preferences and expectations [1]. However, their limitations, such as their susceptibility to generate inaccurate information, difficulty understanding complex contexts, and challenges in reasoning continue to restrict their effectiveness in providing reliable and meaningful interactions [2]. Therefore, creating an agent that can keep coherent, context-aware, and goal-driven conversations remains challenging [3]. Researchers have explored the use of Retrieval Augmented Generation (RAG) as a method to address these limitations, providing a framework for improving the coherence and consistency of conversational responses generated by large language models [4]. RAG has been used in prior works to enhance the coherence and context-awareness of chatbot responses [5].

The field of habit change has seen growing interest in the use of digital assistants, such as chatbots, to support individuals in modifying their behaviors.

Additionally, the use of emotion recognition and empathetic response generation has been explored in the development of conversational agents for mental health applications. For instance, Yu and McGuinness (2024) found that over 70% of participants rated their chatbot highly for conversational quality and supportiveness, suggesting its potential as a valuable tool in providing mental health support through enhanced conversational relevance [6].

This present pilot study focuses on adopting a user-centered design approach to develop and investigate the effectiveness of a GPT-based chatbot, Habit Coach, in facilitating habit change, focusing on the use of Cognitive Behavioral Therapy (CBT) and narrative therapy techniques. We recruited 5 participants to use Habit Coach chatbot for five consecutive days. The Self-Report Habit Index (SRHI) was used to measure the participants' habit strength before and after the intervention. This pilot study builds on this prior work by examining the integration of CBT and narrative therapy techniques within a GPT-based chatbot designed for habit change.

## 2. Literature Review

In the literature, CBT has been used to target and modify many maladaptive behaviors such as substance use, insomnia, unhealthy eating behaviors, and general stress [6]. CBT focuses on identifying and reshaping thought patterns to influence behaviors, using techniques such as self-monitoring, structured goal setting, and skill-building, which encourage individuals to integrate these strategies into their daily lives for behavioral change [6,7]. CBT typically involves a systematic approach to therapy, where sessions are organized around specific goals and techniques aimed at addressing the people's specific issues [6]. This structured framework is particularly suitable for digital adaptation, offering clear steps for the assessment and application of the techniques in a systematic manner.

Narrative therapy, on the other hand, offers an alternative approach that emphasizes the role of personal stories in shaping behavior. By externalizing problems and identifying unique outcomes, narrative therapy enables individuals to see their behaviors as distinct from their identity, promoting the construction of empowering, alternative narratives that align with their desired selves [9,10]. Building on these insights, researchers have explored the potential of chatbots as tools for development of healthy lives, with many studies exploring their role in delivering personalized habit formation and behavior change program [11]. These chatbots have shown promise in areas like increasing physical activity and enhancing user satisfaction, making them useful tools for encouraging positive behavioral changes [12].

Furthermore, AI-powered chatbots in healthcare have emerged as an area of growing interest, offering unique opportunities to influence patient behavior and lifestyle choices [13]. The ability of chatbots to provide accessible, personalized mental health interventions is particularly significant, as they can reduce stigma and offer flexible support that may be difficult to achieve in traditional therapy settings [14]. Non-human counselors can also be considered "free from judgement" by users [15]. The advancement of generative language models, such as ChatGPT, has significantly expanded the possibilities for chatbots in habit formation and behavioral modification. These models allow for more natural and contextualized conversations, allowing users to express themselves openly, and enabling highly personalized interactions, as demonstrated in the Habit Coach explored in this study. Research has shown that AI-based chatbots are effective in promoting behavioral changes, increasing physical activity, and improving adherence to healthy routines [16,17,18]. Yet, most works that investigated the effects of employing a virtual agent to interact with participants to elicit behavior change have constrained participant responses to a set of few options and did not allow the participants to respond openly [19]. Olafsson et al. (2023) reported that participants who can interact with an unconstrained agent had a higher degree of readiness to change their behavior [19]. Despite safety concerns, counseling experts judged these interactions to be safe. Olafsson et al. (2023) also demonstrated that participants who interacted with the agent that delivered motivational interviews and CBT, either constrained or unconstrained, reported higher motivation, confidence, and commitment to behavior change and also, on average, reported that they planned to consume fewer alcoholic drinks in the coming week [19]. They utilized a GPT-2-based LLM and suggested that a newer and more powerful LLM may improve the system [19]. In another study, Xie et al. (2024) designed a chatbot using Motivational Interviewing (MI) techniques to assist users in adopting positive lifestyle changes, particularly for those ambivalent about making changes, such as quitting drug use or improving health [20]. The chatbot, utilizing the DIIR framework, was able to produce collaborative and effective responses, enhancing user motivation and engagement. This highlights the potential of integrating MI techniques into chatbot interventions to support behavior change.

Studies in conversational agents and digital interventions for behavior change highlight the importance of adaptive responses in enhancing user experience and motivation. For instance, adaptive algorithms have been shown to personalize user interactions effectively, improving engagement and facilitating behavior change [23]. Similarly, the design of conversational agents capable of adjusting their responses based on user interactions is critical for maintaining engagement and driving positive outcomes [16]. Additionally, mechanisms such as adaptive feedback have been demonstrated to enhance user motivation and adherence in digital interventions, further supporting the need for tailored responses [19].

However, despite the potential benefits, ethical, legal, and regulatory challenges must be considered, along with the importance of robust natural language processing and machine learning algorithms [21,22]. While chatbots hold promise, the outcomes should be interpreted cautiously, given the moderate to high risk of validity concerns in existing studies [23,24]. As highlighted in previous research, interventions targeting behavior modification require a high degree of contextual adaptation to address individuals' diverse needs, therapeutic techniques, and phases. Achieving such personalization remains a challenge for LLM-based chatbots, which are typically trained on generic user profiles. In this pilot study, we aimed to examine the capabilities and effectiveness of a GPT-based chatbot as a tool to support habit change through the integration of the RAG system and iterative prompting approach. By combining techniques from CBT and narrative therapy, we investigated the chatbot's ability to deliver personalized, context-aware and proactive interactions customized to individual user needs.

# 3. Methodology

We developed and evaluated Habit Coach's effectiveness in supporting users through habit change interventions within a user-centred design framework. A key challenge for the chatbot was to achieve proactive behavior that dynamically adjusts to different therapeutic phases and individual user needs. To address this, we integrated a RAG system to enhance the chatbot's capacity for proactive and context-aware responses. However, our final approach employed an iterative prompt design to customize chatbot's behavior by encoding procedural knowledge in prompts through four main iterative phases. These iterations strengthened the chatbot's ability to deliver personalized interactions in dynamic contexts by improving the system prompt.

## 3.1. Participants

Five participants (Female: 3, Male: 2, Mage = 31.6, SD = 2.65, Range = 27 - 35 years) took part in the study, interacting with the chatbot for a continuous period of 5 days. Full informed consent was obtained from all participants before the beginning of the study. Participants had different backgrounds, the focus being on those who had problematic habits they wished to change; these include eating at night, productivity, and procrastination. The demographic characteristics were diversified on age, professions, and type of problems on habits. Every participant chose a certain habit that they liked to change before the intervention. Over the five-day intervention, participants were directed to use the same chat session every day, enabling the chatbot to rely on the discussion history to ascertain the therapy day. Maintaining a single chat history, the chatbot was able to determine the intervention phase and modify its responses.

## 3.2. Design of the Intervention

This pilot study was structured for participants to interact with the Habit Coach chatbot daily over five consecutive days. While interactive dialogues keep users involved and motivated to implement changes into their lives, explanations assist users comprehend the cognitive and behavioral elements of their habits by fostering self-awareness. Consequently, Habit Coach was designed to include a dynamic switching mechanism that enables it to alternate between interactive dialogues and educational explanations based on user input, need and progress throughout the intervention.

### 3.2.1 Personalization of chatbot behaviors through basic prompts

Initially, we employed a basic prompting approach to personalize the chatbot's behavior across different therapeutic phases, such as encouraging self-reflection, goal setting, and progress tracking. A key challenge, however, was ensuring that the switching mechanism provided educational, context-appropriate guidance while maintaining a natural conversational flow through basic prompting strategy. The chatbot often failed to adjust its tone and content in response to specific user inquiries. Despite efforts to refine the prompts by setting specific rules and boundaries, the chatbot frequently produced lengthy, overwhelming responses that were often repetitive or overly simplistic.

### 3.2.2 Personalization of chatbot behaviors through RAG strategies

To address the issues encountered with the basic prompting approach, we employed a more advanced method, the RAG system, to customize the GPT-based chatbot by leveraging the generative and retrieval capabilities of the RAG framework. The book The Psychology of Habit [25] was uploaded as a source to the system to improve its ability to provide adaptive and relevant interactions. However, additional issues arose. The system's ability to translate declarative knowledge from the book into effective interactive behaviors to modify habits was limited. The chatbot frequently generated complex and verbose responses, asking irrelevant or overly detailed questions that deviated from the prompt's instructions and resulted in long, confusing content.

To create a more structured framework, we then attempted uploading prompt instructions as a document. This approach also proved ineffective, as the chatbot continued to deviate from the intended concise and engaging style. These attempts highlighted the challenges of achieving true adaptability and personalization. Therefore, we concluded that a direct iterative prompt-based approach was more efficient in achieving the desired chatbot behavior.

### 3.2.3 Personalization of chatbot behaviors through procedural knowledge encoded in prompt

We incorporated procedural knowledge directly into the prompt, allowing the chatbot to offer more structured guidance that aligned with specific therapeutic techniques and user needs. Procedural knowledge encoding involved embedding detailed instructions within the system prompt to ensure that the chatbot could provide step-by-step guidance on activities like identifying triggers, setting achievable goals, and practicing self-reflection. This approach allowed the chatbot to offer tailored responses that adapted based on the user's progress, making the interaction feel more intuitive and supportive.

Additionally, encoding procedural knowledge into the prompt allowed the dynamic switching mechanism to work efficiently, enabling the chatbot to switch smoothly between instructing users succinctly and starting introspective dialogues. This design was aimed at helping users when they indicated confusion or struggled with activities. The chatbot was designed, for instance, to go into an explanatory mode when users appeared stuck or answered with "I don't know," offering thorough direction and examples before going back to interactive inquiry. This rapid adaptation allowed chatbot's output to stay relevant, encouraging, and aligned with the user's progress.

### 3.3. Procedural knowledge Prompt Design Iterations

The refinement process of the prompt (see Table 2) provided to the system is outlined as follows.
**First Version:** In the first version, the prompt was designed to engage users in reflective dialogue, emphasizing self-discovery. The chatbot aimed to encourage users to identify their own solutions, fostering a user-driven and open-ended process.

    Example of Interaction:
    Chatbot: "What triggers this habit?"
    User: "I'm not sure. I just snack without thinking sometimes."
    Chatbot: "Can you think of a specific moment when this happens?"
    User: "Not really. Maybe when I'm tired."
    Chatbot: "Okay, tell me more about that."

However, in the initial iteration of the prompt, the chatbot often exhibited a passive role during interactions, relying on users to pose detailed questions or independently lead conversation.

Moreover, it failed to provide actionable suggestions. Instead of offering cues, such as "Do you notice any particular time of day or feeling, like stress or boredom, that leads to snacking?", the chatbot passively relied on the user to elaborate without support. This behavior can result in a loop of vague and unproductive responses, leaving the user feeling disengaged.

**Second Version:** In contrast, a second version was developed to make chatbot more proactive and adopt a structured interaction style by providing step-by-step guidance and giving clear instructions on setting goals and creating action plans (see Table 1). A key improvement between the first and second versions lies in the goal-setting approach. The second prompt iteration enabled the chatbot to adopt a more structured approach, offering specific suggestions and actively assisting users in developing a concrete plan for habit change.

Example of Interaction:
Chatbot: "If I feel stressed, then I will take a walk. Now, create your own If-Then plan."
User: "If I feel stressed, then I will listen to music instead of snacking."

However, this version leaned heavily on a directive approach, which influenced the nature of user interactions. At times, the chatbot responded too rigorously, following exactly its recommendations without changing to reflect user comments. This inflexibility risked disengaging users, as their specific needs or challenges were occasionally overlooked. Additionally, the chatbot's focus on providing solutions sometimes limited users' opportunities to reflect on their own insights, reducing their engagement and sense of responsibility in the habit-change process.

Example of Interaction:
Chatbot: "Can you describe what triggers this habit?"
User: "I usually get bored or stressed, and that leads me to snack late at night."
Chatbot: "What reward do you get from this habit?"
User: "It helps me relax for a bit, but I feel guilty afterward."

**Table 1**
Chatbot Interaction Steps for Habit Change

| Step | Description | Example |
|---|---|---|
| Greeting and Introduction | The chatbot introduces itself and sets the tone for the interaction. | Hi there! I'm your coach today. Let's talk about a habit you'd like to change. |
| Understanding the Habit | The chatbot prompts the user to describe the habit and its triggers. | What specific habit would you like to change, and what triggers it? |
| Identifying Routine and Reward | The chatbot asks about the routine the user follows, and the reward received. | What do you usually do when this habit kicks in? How do you feel afterward? |
| Setting Goals | The chatbot suggests specific, achievable goals for the user. | Could you replace chips with fruit for late-night snacks? |
| Making a Plan | The chatbot helps the user outline a detailed action plan. | When you feel stressed, could you take a walk instead of snacking? |
| Tracking Progress | The chatbot encourages tracking habit change using a journal or app. | How would you like to track your progress? A checklist or a journal? |

**Table 1 (Continued)**

| | | |
|---|---|---|
| Overcoming Obstacles | The chatbot discusses potential obstacles and strategies to overcome them. | What challenges do you anticipate, and how will you deal with them? |
| Regular Check-ins | The chatbot sets up regular check-ins to review progress and adjust the plan. | How often would you like to check in to review your progress? |
| Incorporating Support Systems | The chatbot encourages the user to identify support systems to help. | Who can you ask for support as you work on this habit? |

**Third Version:** In the third version of the prompt, we specifically focused on utilizing GPT's creation feature to enforce a concise dialogue structure by explicitly instructing the chatbot to limit responses to no more than 30 words. This adjustment aimed to enhance the brevity and coherence of interactions while ensuring that the responses remained interactive and on-topic. The modification effectively led to shorter, more targeted responses, thereby aligning the chatbot's behavior with the intended brief interaction model. This improvement addressed prior issues of overly lengthy or irrelevant responses.

Additionally, the third version of prompt iteration focused on including a dynamic switching mechanism that let the chatbot alternately choose between delivering educational instructions and participating in introspective and interactive conversation. This architecture helped adapt chatbot's answers depending on specific user input, while it still preserved brevity and relevancy.

**Final Version:** Last iteration focused on changing chatbots output design to follow a disciplined framework meant for a five-day habit modification intervention by including procedural knowledge on narrative therapy and CBT techniques straight into the prompts. These techniques included spotting and changing behaviors such as visualization success, building powerful stories, and using "if-then" strategies.

Example of Interaction:
Chatbot: "If I feel stressed, then I will take a walk. Now, create your own If-Then plan."
User: "If I feel stressed, then I will listen to music instead of snacking."

We also incorporated structured daily remarks into the prompt to provide users with clear conclusions at the end of each day. These summaries aimed at inspiring users to keep on their path by highlighting their own development.

Example of Conclusion:
Chatbot: "Great job today! Reflect on what you've accomplished. Tomorrow, we'll dive into new techniques to help you stay on track."

By including procedural knowledge straight into the prompts and aligning each exercise with user-specific goals, the last iteration tackled previous issues such as verbosity and lack of proactivity and personalizing.

**Table 2**
Iterative process of the system prompt

| Version | Key Features | Issues Encountered | Solutions Implemented |
| --- | --- | --- | --- |
| 1st | -Providing specific recommendations for habit change | - Passive chatbot<br>- Limited structured advice leading to repetitive or unproductive interactions | - Transitioned to a more directive approach with clear instructions and structured guidance. |
| 2nd | -Introducing goal-setting and action-oriented prompts | -Too structured and impersonal responses<br>-Directive responses and no self-reflection approach | - Balanced directive prompts with space for user-driven input and introspection. |
| 3rd | - Improving response brevity and relevance<br>- Focusing on dynamic switching between interaction and explanation | - Occasional verbosity when switching from explanation to interaction<br>- Challenges in maintaining natural conversational flow | - Further refined prompts to ensure seamless transitions between explanation and interaction. |
| 4th | -Integrating procedural knowledge into prompt<br>-Structured conclusions<br>- Balanced personalized advice with generalizable techniques | - None significant; major issues addressed by iterations | - Finalized prompt structure to match user-specific habits and goals while maintaining concise, engaging responses. |

### 3.4. Description of the final Chatbot Prompt

The chatbot's interactions with users were guided by a structured set of cues and techniques embedded in the comprehensive final system prompt (see Appendix A). Instead of making daily prompt adjustments, a single prompt was utilized to encompass instructions for all five days. The stages from days 1 to 5 for habit change were designed in accordance with the principles of CBT and narrative therapy.

**Day 1:** Identifying and Reframing Habits

Objective: The primary goal of Day 1 is to encourage participants to identify and re-evaluate their daily behaviors, particularly those that are detrimental. Technique: The intervention begins with the externalization of habits, a strategy rooted in Cognitive Behavioral Therapy and Narrative Therapy. Participants are urged to articulate that their behaviors are influenced by external factors. A participant may assert, "Stress drives me to engage in overeating." This reframing allows individuals to detach from their acts and engage in thoughtful discourse. By recognizing that their habits are not solely indicative of their identity but responses to external stimuli, participants can begin to explore alternative coping mechanisms that do not depend on the problematic behavior [26,27].

**Day 2:** Recall Positive Occasions

Objective: The second day focuses on recalling instances where individuals successfully controlled their desires, hence enhancing their self-efficacy. Technique: The chatbot prompts users to reflect on situations where they resisted their actions. Questions may include, "Can you recall a moment when you felt the urge to engage in your habit but chose not to?" What characterized the moment? This reflective activity highlights the participants' capacity for self-regulation and resilience. By identifying the factors that contributed to their success, participants may enhance their confidence in their ability to resist future urges [28].

**Day 3:** Visualization

Objective: The aim of Day 3 is to engage participants in imagining their success in modifying their habits. Tool: Guided imagery is employed as a cognitive-behavioral therapy (CBT) technique, encouraging participants to visualize specific scenarios in which they make positive decisions. For instance, individuals may visualize selecting a nutritious meal instead of yielding to unhealthy snacks. This cognitive representation of success enhances motivation and facilitates the development of new neural pathways associated with positive behavioral transformation [29].

**Day 4:** Formulating the Counter-Narrative

Objective: The fourth day encourages participants to construct an alternative narrative on their actions, so augmenting their sense of agency and empowerment. Technique: Participants are directed to create a modified narrative concerning their behaviors using Narrative Therapy. Individuals are encouraged to adopt the perspective of someone actively engaging in better behaviors, as seen by the assertion, "I manage stress through constructive activities." This reinterpretation of their self-concept allows participants to view themselves as capable of transformation, so enhancing their progress and commitment to developing better behaviors [30].

**Day 5:** Introduction of "If-Then" Strategies

Objective: The final day focuses on developing specific implementation strategies to address potential triggers in the future. The chatbot assists users in formulating "If-Then" plans, a strategy derived from CBT. Participants identify specific triggers associated with their habits and develop actionable responses. For example, "Should I encounter stress, I will take a walk." This proactive approach allows participants to anticipate issues and equips them with targeted methods for regulating their responses to triggers, so increasing the likelihood of successful habit alteration [31].

## 4. Results

### 4.1. Habit Strength

The strength of the targeted habit was assessed by adapting the 12-item Self-Report Habit Index (SRHI) to reflect the specific habit each user aimed to change [32]. Measurements were taken both before and after the intervention. Participants rated their responses on a 7-point scale, ranging from 'Strongly Disagree' (1) to 'Strongly Agree' (7). (e.g., "Drinking coffee is something I would find hard not to do.").

We used Jamovi (V 2.4.14.0) to analyze the data. A paired samples t-test was conducted to compare self-reported habit strength before (M = 52.20, SD = 3.35) and after the intervention (M = 29.20, SD = 2.77). The results of our pilot study showed that the intervention was effective in significantly reducing the strength of participants' habits; $t(4) = 25.71$, $p < .001$.

**Table 3**
SRHI Results of Participants Before and After Intervention

| Participant | Target Habit | Pre-Intervention SRHI | Post-Intervention SRHI | Change in Habit |
|---|---|---|---|---|
| Participant 1 | Late-night snacking | 48 | 28 | 41.67% |
| Participant 2 | Procrastination | 52 | 30 | 42.86% |
| Participant 3 | Social Media Overuse | 56 | 32 | 42.86% |
| Participant 4 | Bedtime Procrastination | 50 | 25 | 50% |
| Participant 5 | Coffee Intake | 55 | 31 | 43.64% |

### 4.2. Feedback

Additionally, participants were asked to provide feedback on their experiences with the chatbot considering their target habits after the intervention (see Table 3). Their responses provided more insights into how different aspects of the intervention, including the contribution of CBT and narrative therapy techniques, influenced their behaviors. The following summarizes key themes and individual experiences reported by participants:

- **Participant 1:** Reported reduced snacking at night and a more regulated evening eating routine. The chatbot's CBT methods helped increase awareness of emotional triggers like boredom.
- **Participant 2:** Experienced a significant reduction in procrastination and an increase in the completion of assignments. Narrative therapy techniques assisted in reframing procrastination as something that could be controlled.
- **Participant 3:** Saw an increase in productivity, attributing this to visualization techniques and time management strategies suggested by the chatbot.
- **Participant 4:** Reported an increase in energy levels and established a regular morning routine.
- **Participant 5:** Reduced coffee intake, which improved energy and sleep. The chatbot provided personalized suggestions for healthier substitutes, like herbal teas.

The findings revealed a reduction in habit strength in negative habits following the use of Habit Coach, suggesting that integrating CBT and narrative therapy techniques within conversational agents is a promising approach for supporting behavioral change.

## 5. Conclusion

The findings of this study suggest that using a GPT-based chatbot to deliver habit change interventions can be an effective and scalable approach. The chatbot successfully incorporated CBT and narrative therapy techniques to support participants in developing and strengthening healthy habits. Our initial attempts at integrating a RAG system and later incorporating a document with prompts revealed several challenges. While the chatbot was provided with declarative knowledge from therapy techniques through reference materials, it struggled to consistently transform this static information into dynamic and contextually appropriate interactions. As a result, we shifted our focus to making iterative adjustments to the prompts in order to enhance the chatbot's interaction strategies. This approach gradually improved the chatbot's ability to engage in more personalized and context-driven conversations, effectively motivating habit change.

There are several limitations that must be considered. The limited sample size of five participants limits the broader applicability of the results. Future research should not only focus on refining the chatbot's conversational abilities but also include larger and more diverse participant groups to strengthen the validity and generalizability of the findings. Additionally, it was not possible to conduct a detailed quantitative analysis of the chatbot's adaptability to varying user conditions and requests with a small sample size. While LLM-based chatbots are recognized for their flexibility and natural language understanding, they also exhibit unique types of failures that require ad hoc tools for systematic investigation [33]. Addressing these challenges often requires substantial annotation efforts and in-depth analysis.

Secondly, the study did not include a control condition, which limits the ability to fully isolate the impact of the intervention. As this was a pilot study aimed at exploring whether a RAG system could be employed to refine the chatbot's conversational abilities and personalize interventions targeting habit change, the focus was primarily on the design process of the chatbot. The exploratory nature of the study and the early-stage testing of the system focused on feasibility rather than establishing efficacy. Nevertheless, future research would benefit from including a control group to strengthen causal inferences about the chatbot's effectiveness.

Moreover, the use of a self-report scale to measure habit strength, while convenient, introduces potential biases such as social desirability [34] or inaccurate self-assessment. Future research could complement self-report measures with behavioral assessments, such as tracking actual behavioral changes over time (e.g., using wearable devices, activity logs, or direct observation) to provide more objective data on habit modification.

Lastly, while the chatbot was tested over five consecutive days, the results might not fully represent behavior change in the long term or user engagement over more extended periods. This short duration limits the ability to assess whether habit changes are sustainable in the long run. Future studies should investigate the chatbot's effectiveness over a longer intervention period to assess the persistence of habit modification. Additionally, using the same chat session for 5 consecutive days streamlined the design and interaction process but depended on the presumption that participants would follow exactly the instructions not to start fresh chat sessions. Although this approach worked well for this pilot study, future implementations might include automatic session tracking or journaling systems to lower dependence on user behavior and improve system adaptability.

## Acknowledgement

This publication has emanated from research supported in part by a grant from Science Foundation Ireland under Grant number 18/CRT/6183. For the purpose of Open Access, the author has applied a CC BY public copyright license to any Author Accepted Manuscript version arising from this submission.

## Appendix A

Your goal is to coach the user on how to change their bad habits into good ones. Interact with the user by asking short questions, giving short feedback, and providing instructions in bulleted steps. Use evidence-based strategies explained specifically in each day. Switch to explaining whenever the user doesn't have ideas, then switch back to the interaction after explaining. Never ask more than one question in each response. Always give examples that are not in user's field of habit. For example, if they talk about smoking, you give example about workout, let them write their own ideas. Give them an exercise each day. Do a conclusion at the end of each day. Do exactly this way, don't act creatively.

Day 1: Introduction to Habit Formation: Identify the habit you want to change and the perceived underlying reason. Example: "I smoke when stressed." Your Turn: (Write your habit here). Step 2: Reframe the Habit: Reframe the habit to externalize it. Example: "Stress is affecting me to smoke " Your Turn: (Reframe your habit here). Step 3: Have a dialog with user, ask it if they can control themselves. Conclusion: Great start! Reflect on what you've written today. Tomorrow, we'll look at positive instances when you resisted this habit. See you then!

Day 2: Recall Positive Instances Objective: Identify times when you resisted or overcame the habit. Step 1: Recall Positive Instances: Think of times you resisted the habit. Example: "I felt stressed but took a walk instead of smoking." Your Turn: (Write your instances here) Step 2: Analyze: What Was Different Reflect on what made these times different. Example: "I chose to walk because I enjoy fresh air." Your Turn: (Analyze what was different here) Conclusion: Well done! Recognizing these moments is crucial. Tomorrow, we'll visualize your success. See you then!

Day 3: Visualization Objective: Use visualization to reinforce new habits. Step 1: Find a Quiet Space: Choose a quiet place where you won't be disturbed. Your Turn: (Find your space) Step 2: Visualize Success: Visualize yourself successfully changing the habit. Example: Imagine choosing a healthy behavior over the old habit. Your Turn: (Visualize your success) Conclusion: Excellent! Visualization helps reinforce new habits. Tomorrow, we'll write a new, empowering narrative. See you then!

Day 4: Write a New Narrative Objective: Create an empowering narrative about your new habit. Step 1: Write a New Narrative: Write a positive statement about your ability to change. Example: "I manage stress through healthy activities." Your Turn: (Write your narrative here) Conclusion: Fantastic! Your new narrative will guide your behavior. Tomorrow, we'll implement "If-Then" plans to solidify your habit change. See you then!

Day 5: Implement "If-Then" Plans Objective: Form specific implementation intentions. Step 1: Create an If-Then Plan: Link a trigger with new, healthier behavior. Example: "If I feel stressed, then I will take a walk." Your Turn: (Write your If-Then plan here) Conclusion: Great job! You've set a solid foundation for changing your habit. Keep practicing these steps, and remember, every small effort counts towards your overall success. Continue your journey and celebrate your progress!